# Design of Muon Campus full flow purifier for varying operational conditions and horizontal shipping

**J. Subedi, T. Tope, B. Hansen, Y. Jia, J. Makara, J. Tillman, Z. Tang**

Fermi National Accelerator Laboratory, Batavia, IL 60510

Email: jsubedi@fnal.gov

**Abstract**. Constant ingress of impurities in Muon Campus g-2 experiment at Fermilab has resulted in reduction of efficiency of cryogenic expanders and occasional undesired downtime to flush the impurities. Due to insufficiency of current 60 g/s mobile purifier, a full flow purifier is designed to be used in Muon Campus which purifies 240 g/s of Helium throughput of 4 compressors through charcoal bed at 80 K and returns ambient Helium back to the system. The purifier is designed to be operated near liquid Nitrogen temperature during cold operations and up to 400 K during regeneration. Both warm and cold operational range of the purifier has required use of appropriate clearances in design due to expansion and contraction. The vessel of around 16 ft height which is designed to be operated vertically is to be shipped horizontally. The asymmetrical position of heavy stainless steel heat exchanger in the purifier support frame and 5g vertical load design consideration for shipping has required use of shipping supports and heat exchanger rotational stops to comply with design requirements. FEA of purifier system is performed in cold, warm and shipping cases to verify that the purifier satisfies the design requirements.

## 1. Background

The Muon Campus g-2 experiment requires stable helium supply at 5 K to the superconducting magnet. After a short period of operations, the pressure drop across the magnet's flow supply valve increases due to impurities, resulting in reduced excess refrigerator capacity concurrently with drop in refrigerator expander efficiency. This requires periodic powering down of the magnet to allow for helium valve "flush" to remove accumulated contamination at the valve as well as warming up the refrigerator expander above 80 K to release contamination. A "mobile" purifier is used for this periodic process but only at 10% of total refrigerator flow such that the purification process is extended. After long duration of running, this purification procedure is required almost every two weeks to restore proper liquid helium supply to the magnet. This shows requirement for full flow purifier which can purify impurities from entire Muon Campus refrigerator system and mitigate existing impurities issue and associated experiment downtime.

## 2. Sizing of a purifier

The sizing of purifier is based on assumption that there is a small but constant ingress of impurity in the form of air consisting mostly of Nitrogen entering in the system. This impurity is assumed to be 2 ppm impurity of Nitrogen which is to be adsorbed in purifier system. The excess adsorption energy of activated Carbon used as adsorbent in purifier is calculated based on below formula of Polanyi Potential Theory [1,2].

$$\epsilon_{ij_{eq}} = RT \ln(P_s/p)$$

where,
R = Gas Constant for Nitrogen
T = Temperature of adsorber bed
$P_s$ = Saturation pressure
p = Partial pressure of Nitrogen

The partial pressure of Nitrogen is assumed to be directly proportional to the concentration of Nitrogen in Helium. Thus, partial pressure of Nitrogen is determined by multiplying impurity concentration ratio with total pressure of Helium system [2].

The saturation pressure of a Nitrogen is determined for a given carbon bed temperature T. Once, excess adsorption energy is determined, the amount of Nitrogen adsorbed per weight of Carbon is determined from Characteristic curve for nitrogen on PCB carbon [3]. Total amount of Nitrogen that needs to be adsorbed over a period of purifier online time can be determined by calculating amount of Nitrogen that gets accumulated based on inlet impurity at given flow rate. Total amount of Charcoal required is thus determined by below formula.

$$\text{Total charcoal weight required} = \frac{\text{Total Nitrogen that needs to be adsorbed}}{\text{Nitrogen adsorbed per weight of charcoal}}$$

The charcoal historically used in the purifier is Sorbonorit ® B 4 activated carbon. Based on the weight and density of the charcoal; total volume occupied by the charcoal is calculated. The adsorber volume is thus equivalent to the total volume of charcoal required for adsorption.

Based on the operating conditions at Muon Campus, as tabulated in Table 1 the charcoal volume of Sorbonorit ® B 4 required was calculated as 34 ft$^3$. Based on L/D ratio of 6 which has been historically used for other purifiers, the final dimension of adsorber vessel came out to be 24-inch diameter and 144-inch length. The sizing of Nitrogen jacket was based on volume capacity of at least 80 gallons. The vacuum vessel was designed based on size of the adsorber vessel, heat exchanger and support needed for internal vessels.

| Input Parameters | Value | Units |
|---|---|---|
| Impurity adsorption level | 2 | ppm |
| Adsorber bed pressure | 20 | atm |
| Helium mass flowrate | 240 | g/s |
| Adsorber online time | 180 | days |
| L/D ratio | 6 | - |

Table 1: Inlet and input parameters for adsorber sizing

| Result | Value | Units |
|---|---|---|
| Charcoal volume | 34 | ft^3 |
| Adsorber length | 144 | inch |
| Adsorber diameter | 24 | inch |

Table 2: Final sizing of adsorber vessel based on inlet and input parameters.

## 3. 3D model of a purifier

3D model of purifier was developed by Ability Engineering Technology Inc. based on the Engineering Specification Document and conceptual 3D model from Fermilab. The purifier is designed to be shipped in horizontal orientation. The piping and components on vacuum vessel are oriented such that area contacting the lower surface is clear of these components to facilitate horizontal shipping. The reliefs are located such that they are easily accessible for maintenance and replacement. There are 7 permanent shipping support pads which support the internal frame supporting the adsorber vessel and heat exchanger and prevent large displacements.

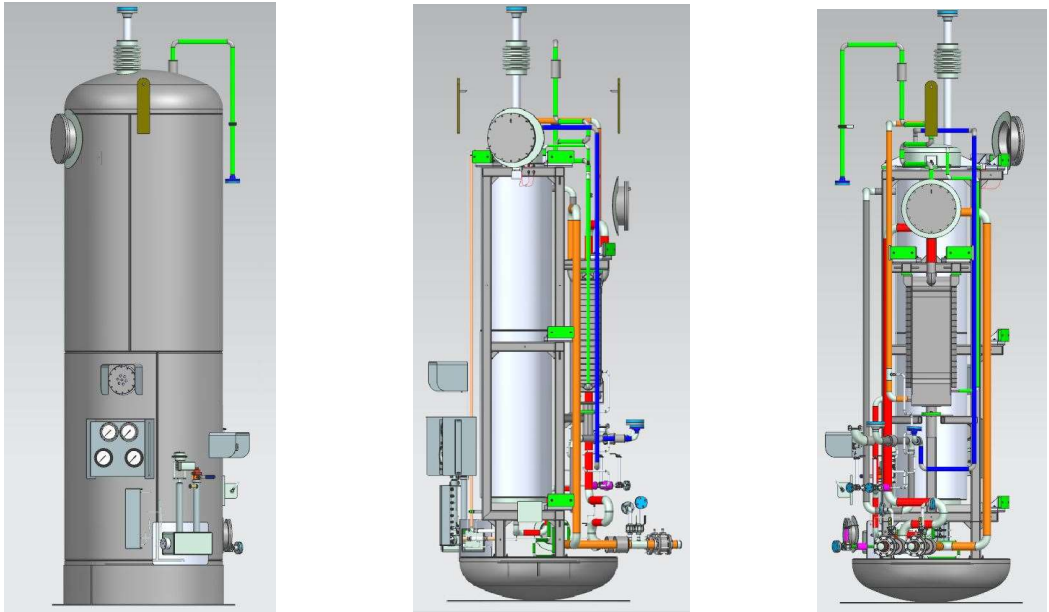

Figure 1: 3D model of purifier showing full assembly and internal view (Image courtesy: Ability Engineering Technology Inc.)

### 4. FEA of Purifier
Finite Element Analysis was performed for 3D model for operational and shipping cases in collaboration with Ability Engineering Technology Inc. and the results were used to determine high stress and displacement locations to modify support for the purifier.

The purifier is designed to operate at 80 K and undergo regeneration at temperature up to 400 K. Thus, finite element analysis was performed in ANSYS for both warm and cold conditions to determine displacement and stresses in the components of the purifier. For warm case, two different sub-cases were considered; one where heat exchanger remains cold during regeneration and another where heat exchanger warms up along with adsorber vessel during regeneration. For both these warm cases stress and displacements were similar. The maximum allowable stress was established as 138 MPa for 304 L stainless steel. Stresses in operational cases were below allowable stress except lower part of frame where there are localized high stresses. Further plastic-elastic analysis was performed to show that the localized high stresses were acceptable and did not cause plastic collapse.

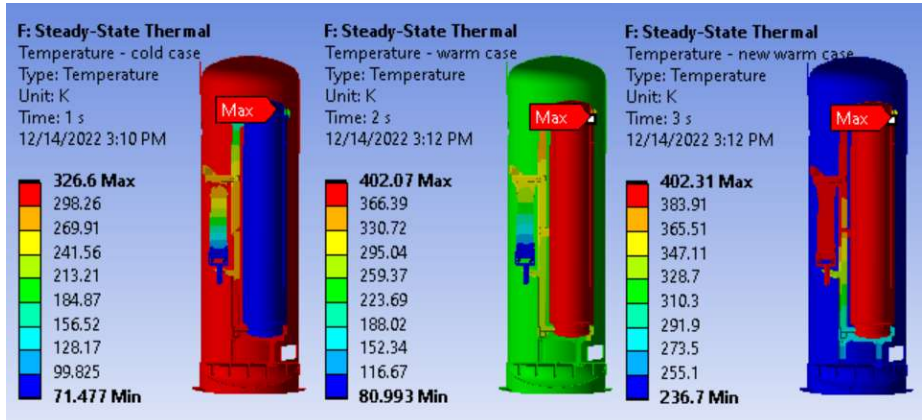

Figure 2: Temperature distribution for cold and 2 warm operational cases of purifier

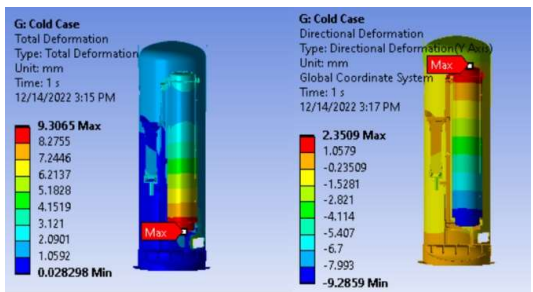

Figure 3: Total and directional deformation for for cold operational case of purifier

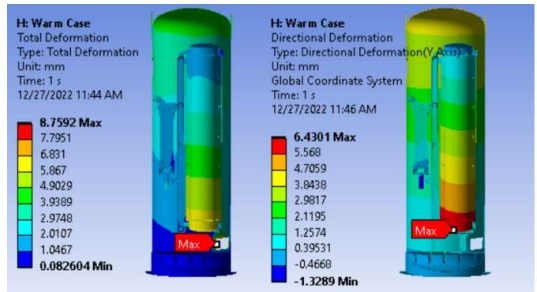

Figure 4: Total and directional deformation warm operational case of purifier

For horizontal shipping of purifier, the maximum acceleration effect is assumed as 2g in longitudinal and lateral direction and 5g in vertical direction. Highest stress and displacements were seen in the 5g case in the same direction as weight. In addition, high local stresses were seen in frame at certain locations. Elastic-plastic analysis was done for most severe shipping load case.

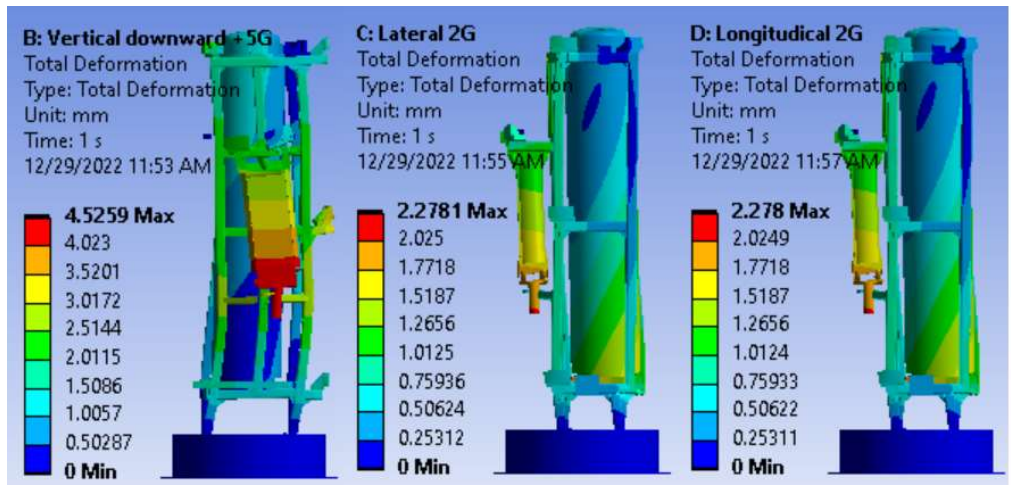

Figure 5 : Total deformation in vertical 5g, lateral 2g and logntudinal 2g cases without outer vessel

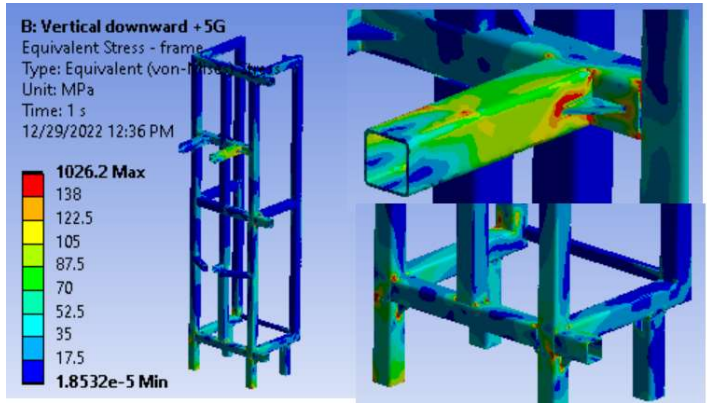

Figure 6: Equivalent stress of purifier frame in 5g vertical downward case

The load in an elastic-plastic analysis is 3 times that of design load. The load is applied in 3 load steps: 5g, 10g and 15g. The displacement vs load step plot is shown in Figure 7. As converged solution is achieved, it means there is no global plastic collapse [4].

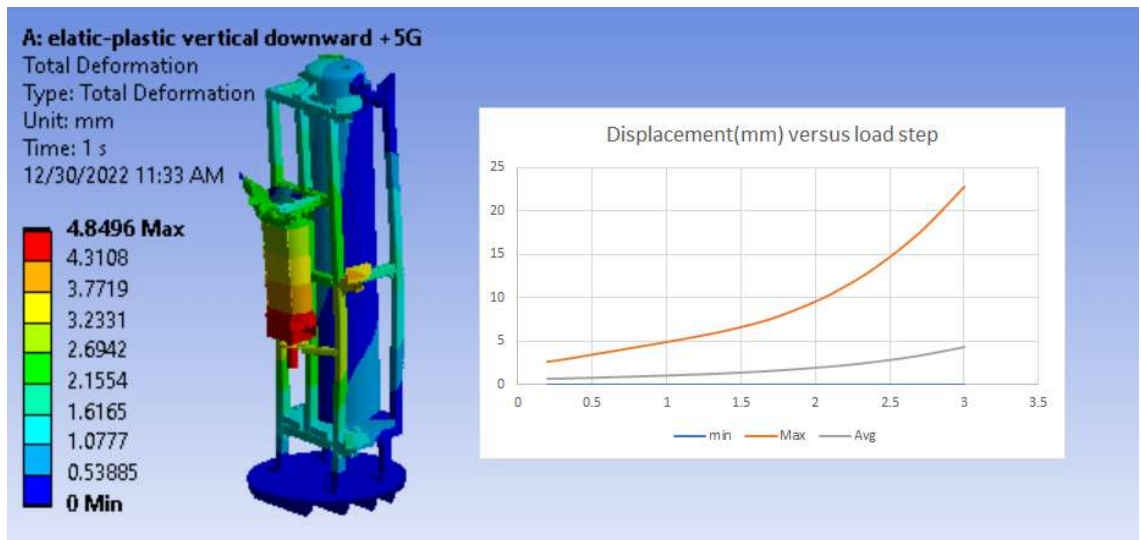

Figure 7: Minimum, average and maximum displacement plots for 3 load steps: 5g, 10g and 15g

Elastic plastic analysis-local strain limit was determined according to ASME BPVC VIII-2 section 5.3 protection against local failure. As per this analysis, component is acceptable for specified load case if sum of forming strain and equivalent plastic strain is lower than limiting triaxial strain. ANSYS has a tool to calculate the strain ratios of the sum of forming and equivalent plastic strain to limiting triaxial strain. This maximum local strain ratio or local failure expression is less than 1 as shown in Figure 8 indicating no danger of local failure in the high stress purifier locations.

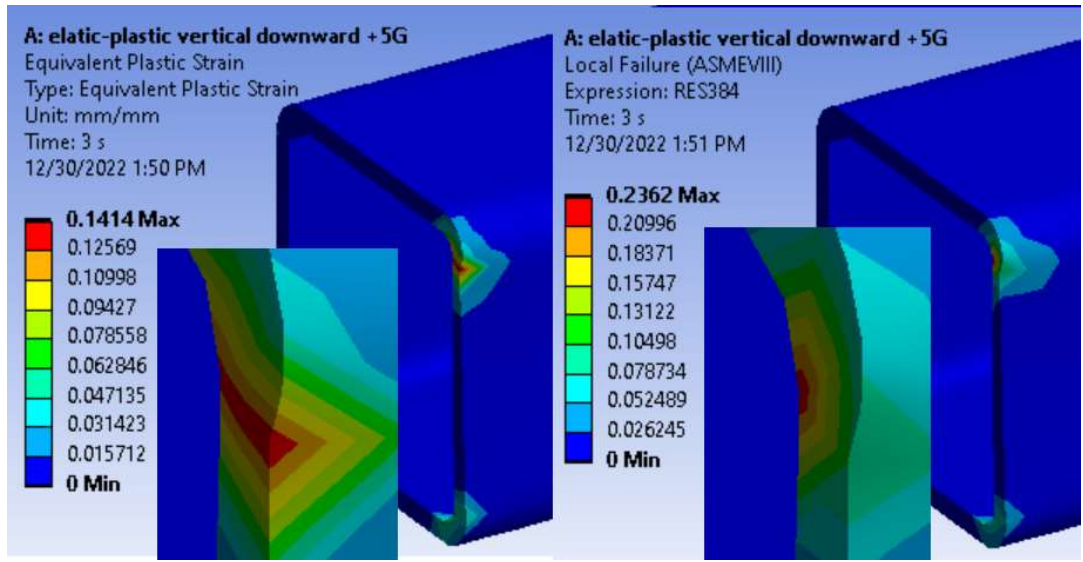

Figure 8: Equivalent plastic strain for 5g vertical downward case and elastic-plastic local strain ratio showing maximum value lower than 1.

Thus, with the help of iterative finite element analysis through ANSYS and modifications, results were achieved where the displacements and equivalent stresses within the purifier were in acceptable range for operational and shipping cases.

### 5. Modifications based on FEA

Previous iteration of FEA analysis showed that offset location of stainless-steel heat exchanger weighing over 1500 lbs generated very high stress and displacement in the internal frame support of the purifier for 5g shipping case. FEA showed the requirement for additional support of purifier frame for 5g shipping case. The heat exchanger itself had high displacement due to its location as shown in FEA analysis. Thus, two modifications were made to resolve the issue. First, heat exchanger stops were designed to be welded on to vacuum vessel and act on heat exchanger shipping support pads which would restrict movement of heat exchanger during 5g shipping case resulting in reduction of equivalent stress.

The requirement of additional support for support frame was fulfilled by using a temporary shipping support bar bolted between vacuum vessel and support frame. This aluminum bar would only be used during shipping and uninstalled through a manhole once the purifier is vertically installed in the location.

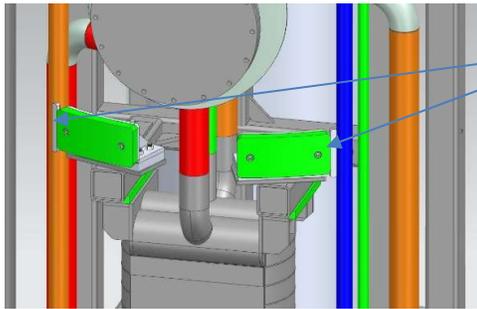
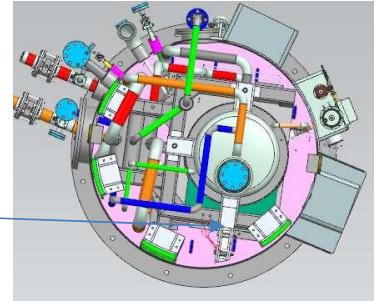

Figure 9: 3D model showing heat exchanger stops on shipping support pads.

Figure 10: 3D model showing temporary shipping support bar.

## 6. Conclusion

Sizing of purifier was done based on inlet conditions and input parameters for Muon campus. Finite Element Analysis was used iteratively to resolve issues of high equivalent stresses and displacements during operating and shipping conditions by addition of supports.

## 7. References

[1] Polanyi, M. "Verh. deut. physik." *Ges.* 18 (1916): 55.
[2] Wright, Mathew. DESIGN AND DEVELOPMENT OF A HELIUM PURIFIER. United States.
[3] Manes, M. I. L. T. O. N., & Grant, R. J. (1963). Calculation methods for the design of regenerative cryosorption pumping systems. In 1963 Transactions of the Tenth National Vacuum Symposium of the American Vacuum Society, edited by George H. Bancroft, The MacMillan Company, NY (p. 122).
[4] ASME BPVC VIII-2 2017

## Acknowledgments

This manuscript has been authored by Fermi Research Alliance, LLC under Contract No. DE-AC02-07CH11359 with the U.S. Department of Energy, Office of Science, Office of High Energy Physics. The authors wish to recognize the dedication and skills of APS-TD/Cryogenics technical staff involved in the installation and commissioning of this system.